\documentclass[
    ,final            
  ]
  {aipproc}
\layoutstyle{8x11single}
\usepackage[mathcal]{eucal}
 \usepackage{amsmath}

\begin{document}
\title{Quantum Deformations of Einstein's Relativistic Symmetries
}


\author{Jerzy Lukierski}{
  address={Institute for Theoretical Physics\\ University of Wroc\l aw,
  pl. M. Borna 9, 50-204 Wroc\l aw, Poland}}

\begin{abstract}
We shall outline two ways of introducing the modification
of Einstein's relativistic symmetries of special relativity theory
- the Poincar\'{e} symmetries. The most complete way of
introducing the modifications is via the noncocommutative
Hopf-algebraic structure describing quantum symmetries.
Two types of quantum relativistic symmetries are described,
one with constant commutator
of quantum Minkowski space coordinates ($\theta_{\mu\nu}$-deformation) and
second with Lie-algebraic structure of quantum space-time, introducing  so-called
$\kappa$-deformation.
The third
fundamental constant of Nature - fundamental mass $\kappa$ or
length $\lambda$ - appears naturally in proposed quantum relativistic symmetry scheme.
The deformed Minkowski space is described
as the representation space
 (Hopf-module) of deformed Poincar\'{e} algebra. Some possible
perspectives of quantum-deformed relativistic symmetries will be
outlined.
\end{abstract}

\maketitle


\section{Introduction}
It is well-known that the nonrelativistic symmetries are described
by the Galilei group, with the Galilean boosts  relating
dynamically equivalent frames which move with relative constant
velocity $\overrightarrow{v}=(v_1,v_2,v_3)$
\begin{equation}
\label{konlu1}
x'_i = x_i + v_i t' \qquad t'=t \, .
\end{equation}
The velocity values  $v=|\overrightarrow{v}|$ are not bounded and three
boost generators $K_i$ ($i=1,2,3$) commute
\begin{equation}
\label{konlu2}
[K_i, K_j] = 0 \, ,
\end{equation}
Einstein's equivalence of relativistic frames is described by the
following modification of \eqref{konlu1} (we choose for simplicity
$\overrightarrow{v}=(0,0,v)$,
\begin{eqnarray}
\label{konlu3}
x'_1 =  x_1     \qquad x'_2 = x_2\, , \qquad   &&
\cr\cr
x'_3 =\cosh \alpha x_3 + \sinh \alpha x_0  &&\quad x_0 = ct\, ,
\cr\cr
x'_0 = \sinh \alpha x_3 +\cosh \alpha x_0 &&\quad \tan \alpha = \frac{v}{c}\, .
\end{eqnarray}
The Lorentz boosts $N_i$ generating  pseudo-orthogonal
rotations in the  Lobachevsky planes
$(x_i,x_0)$ are described by the deformation of the Abelian algebra \eqref{konlu2}

\begin{equation}
\label{konlu4}
[N_i, N_j] = \frac{1}{c^2}\, M_{ij} = \frac{1}{c^2} \ \epsilon_{ijk} \, M_k \, ,
\end{equation}
where $M_{ij}= - M_{ji}$ generate the space rotations in ($i,j$) plane
 ($i,j=1,2,3$).

 The relativistic transformation \eqref{konlu3} should be applied to moving frames if
  the relative velocity ration $\frac{v}{c}$ is not negligible.
  Further, from the invariance of Maxwell equations under Poincar\'{e}
  symmetries follows that $c$ should be interpreted as the velocity of
  electromagnetic waves (light velocity).

  The Poincar\'{e} symmetries of special relativity theory are described by
   10 generators $I_A=(M_{\mu \nu}, P_\mu; \mu,\nu=0,1,2,3)$ which satisfy
   the Poincar\'{e} Lie algebra (to compare with \eqref{konlu4} we
   should put $M_{i0}=c N_i$)
   \begin{eqnarray}
   \label{konlu5}
&&   [M_{\mu\nu}, M_{\rho \tau} ] =   \eta_{\mu\rho} \, M_{\nu\tau} -
\eta_{\nu\rho} \, M_{\mu\tau} + \eta_{\nu\tau}\, M_{\mu\rho} -
\eta_{\mu\tau}\, M_{\nu\rho} \, ,
   \cr\cr
 &&  [M_{\mu\nu}, P_{\rho } ] = \eta_{\mu\rho} \, P_\nu - \eta_{\nu\rho}\, P_\mu \, ,
   \cr\cr
  && [P_{\mu}, P_{\nu} ] = 0\, .
   \end{eqnarray}

The Poincar\'{e} algebra was
 considered for almost a century after Einstein's discovery in 1905
 as quite  uncontested way of describing the equivalent space-time
 frames in relativistic elementary particle physics,
  however under the assumption that the
 gravitational effects are negligible. Further in consistency with the
 classical Poincar\'{e} group structure (Abelian translation group!)
  it was
 assumed that the relativistic space-time is described by classical commuting
 Minkowski space-time coordinates $x_\mu$, i.e.

\begin{equation}
\label{konlu6}
[x _\mu, x_\nu] = 0\, .
\end{equation}
In last period however,   the status of the relations
\eqref{konlu5} and \eqref{konlu6}  as describing ultimate geometric description of
microworld physics was challenged.

 Let us recall here some arguments leading to the modification of
Einstein's relativistic symmetries
 and classical
nature of space-time.

i) Due to Einstein's general relativity theory the space-time manifold with
its geometrical structure is a dynamical entity, with its metric (in quantum theory)
 undergoing
 the quantum fluctuations. In particular any measurement of
position due to Heisenberg uncertainty relations leads to
disturbed
 values of the energy density, what affects through
 Einstein's GR equations the values of gravitational field.
If we calculate the fluctuations of gravitational field generated by the
energy needed to measure the distance with accuracy $\Delta x_\mu$ one arrives at the
conclusion, that it is not possible to measure distances which are smaller than
the Planck length $l_P$ ($l_P \simeq 10^{-33}$cm).
 We see therefore that due to quantum gravity effects
the space-time
 from operational point of view ceases to be a
continuous manifold, but becomes a discrete set  of small cells with the Planck
length size $l_P$. Algebraically the noncontinuous structure of "quantum"
Minkowski space can be expressed by the Dopplicher-Haag-Roberts (DHR) relation
\cite{konlu1}.
\begin{equation}
\label{konlu7}
[\hat{x} _\mu, \hat{x}_\nu] = l_P^2 \, \theta_{\mu\nu} \, ,
\qquad  \theta_{\mu\nu}= -  \theta_{\nu\mu} \, ,
\end{equation}
where in simplified version of the model the tensor $ \theta_{\mu\nu}$ is central,
i.e. one can postulate that it takes numerical values.

ii) Other argument comes from the brane world scenario, in which we assume
that the space-time manifold is described by the $D$-brane coordinates \cite{konlu2}.
The $D$-branes describe the location of the end points of strings (in general of fundamental
$p$-branes). If we accept the view that the fundamental super-strings
describing the most elementary objects in Universe are ten-dimensional, the
$D$-branes are coupled not only to the gravitational background $g_{\mu\nu}(x)$
but as well to the nonvanishing antisymmetric tensor field $B_{\mu\nu}(x)$
 (here $\mu,\nu=0,1, \ldots, 9$). It was shown using canonical quantization
 techniques  (see e.g. \cite{konlu3,konlu4})
 that the quantized end point $\hat{x}_\mu$ of the open string after
  first quantization do satisfy
 the following noncommutativity relation
\begin{equation}
\label{konlu8}
[\hat{x} _\mu, \hat{x}_\nu] = \theta_{\mu\nu}
 \, (B(\hat{x})) \, ,
\end{equation}
i.e. $\theta_{\mu\nu}$ is a local function of $B_{\mu\nu}(x)$.
If we live on $D$-brane, the relation \eqref{konlu8} describes the
noncommutative $D=10$ space-time algebra, which after dimensional
reduction $D=10 \to D=4$ implies the quantum nature
of our four-dimensional Minkowski space.

iii) Third source of the space-time noncommutativity is  provided by the spin
degrees of freedom. Already in seventies it was shown \cite{konlu5}
that the space-time coordinates of superparticles satisfy the following
equal time (ET) relation
\begin{equation}
\label{konlu9}
[{x} _\mu, {x}_\nu] = \frac{i}{m^2} \, S_{\mu\nu}\, ,
\end{equation}
where the relativistic spin operator $S_{\mu\nu}$  satisfies Lorentz
algebra relations and can be expressed in terms
of Pauli-Lubanski vector $W_\mu$. Independently, the formula \eqref{konlu9}
was confirmed in twistor theory, with space-time variables as composites of
primary twistor coordinates \cite{konlu6}. It should be added that old Snyder idea of noncommutative
space-time (see e.g. \cite{konlu7})  also relates the noncommutativity with nonvanishing angular momentum.

In this short talk we would like to recall the quantum deformation of Einstein's
symmetry scheme which is consistent with simple models of noncommutative space-time.
Because the space-time coordinates can be also described by translation sector of
Poincar\'{e} group,
 in order
to describe noncommutative space-time one should consider new algebraic relativistic
symmetries with noncommutative symmetry parameters. We shall consider below in Sect. 2
 and  Sect.  3
 two types of such quantum symmetries:
 the canonical $\theta_{\mu\nu}$-deformation Poincar\'{e} symmetries, implying the
relation \eqref{konlu7}
 with the constant values of tensor $\theta_{\mu\nu}$, and
 the so-called $\kappa$-deformed
relativistic symmetries.
These two types of deformations  introduce new fundamental mass parameter $\kappa$
which enters the respective noncommutativity   relations of
Minkowski space coordinates.
 In Sect. 4 we shall
 mention possible prospects of
deformed relativistic symmetries.

It should be pointed out that the validity of quantum relativistic symmetries can
be also interpreted as the particular violation of classical Poincar\'{e} symmetries
 \cite{konlu8a}.
 {} For example if we consider
 the modification of the
classical Poincar\'{e} mass Casimir,  which follows from
the $\kappa$-deformation of Poincar\'{e} algebra \cite{konlu8,konlu9},
 we obtain
\begin{equation}
\label{konlu11}
C_2 = p^2_0 - \overrightarrow{p}^{ 2} \to C^{\kappa}_2 =
4 \kappa^2 \sin^2 \frac{p_0}{2\kappa} -
\overrightarrow{p}^{ 2} =
p^2_0 - \overrightarrow{p}^2 - \frac{1}{\kappa}\, p^3_0 + {\mathcal{O}}(\frac{1}{\kappa^2})\, ,
\end{equation}
The formula (\ref{konlu11}) for $C_2^\kappa$
  can be interpreted in two ways

i) As a result of a  particular violation of classical
 Lorentz invariance by $\frac{1}{\kappa}$ terms

ii) The indication that the classical Lorentz invariance and
 the  Lorentz transformations
should be modified in a way which leads to the modified
 mass Casimir  $C^\kappa_2$ as a new invariant quantity.

The advantage of second point of view lies in quite
 strong limitations on the ways in which the classical Einstein symmetries
  can be
violated. In this talk we assume that the modification of classical
symmetries
 is represented by a Hopf-algebraic structure. Our hope is
 that in similar way as  hundred years ago the
  Einstein symmetries replaced Galilei ones, in
future considerations e.g. due to the quantum gravity effects the
Einstein symmetries will be  modified and replaced by quantum
 relativistic symmetries.

\section{$\boldsymbol{\theta_{\boldsymbol{\boldsymbol\mu\nu}}}$-deformation: An example of Soft  quantum
Deformation of Einstein's relativistic
 symmetries}

Let us expand the rhs of the general noncommutativity
  relation 
\begin{equation}
    \left[\hat{x}_\mu, \hat{x}_\nu\right]
    =\frac{1}{\kappa^2}\theta_{\mu\nu}\left(\kappa\hat{x}\right)\,,
    \label{konlu11a}
\end{equation}
where $\kappa$ is a masslike parameter, as follows:
\renewcommand{\theequation}{\arabic{equation}a}
\setcounter{equation}{11}
\begin{equation}
\label{konlu11aa}
\theta_{\mu\nu}\left(\kappa\hat{x}\right)=\theta_{\mu\nu}^{(0)}+\kappa\theta_{\mu\nu}^{(1)\  \rho}\hat{x}_\rho+\kappa^2\theta_{\mu\nu}^{(2)\ \rho\tau}\hat{x}_\rho\hat{x}_\tau+\dots\,.
\end{equation}
\renewcommand{\theequation}{\arabic{equation}}
\setcounter{equation}{12}
If in the power serie \eqref{konlu11a} only first term is nonvanishing
($\theta_{\mu\nu}(\hat{x}) \equiv \theta^{(0)}_{\mu\nu}$) we shall
call the modified Poincar\'{e} symmetries preserving the relation
\eqref{konlu8} the
canonically deformed or $\theta_{\mu\nu}$ -deformed Poincar\'{e} symmetries. Such
quantum symmetries
 were discovered quite recently \cite{konlu10}--\cite{konlu14}
  and are obtained by the modification
of classical Poincar\'{e}-Hopf algebra only in the Lorentz
coalgebra sector by so-called twist function \cite{konlu15}
\begin{equation}
\mathcal{F}_{\theta }=\exp \,\frac{i}{2\kappa^2}(\,\theta ^{\mu \nu }_{(0)}\,P_{\mu }\wedge P_{\nu
}\,).  \label{konlu12}
\end{equation}
The generators $(M_{\mu\nu}, P_\mu)$ satisfy the classical Poincar\'{e}
algebra and the modified Poincar\'{e} coalgebra looks as follows
\begin{eqnarray}
\Delta_\theta(P_\mu)&=&\Delta_0(P_\mu), \nonumber\\
\Delta _{\theta }(M_{\mu \nu })&
=&\Delta _{0}(M_{\mu \nu })-\frac{1}{\kappa ^{2}}%
\theta ^{\rho \sigma }_{(0)}[(\eta _{\rho \mu }P_{\nu }-\eta _{\rho \nu
}\,P_{\mu })\otimes P_{\sigma }\\
&& + P_{\rho}\otimes (\eta_{\sigma \mu}P_{\nu}-\eta_{\sigma \nu}P_{\mu})]\,,
\nonumber\label{konlu13}
\end{eqnarray}
where ($I_A = (M_{\mu\nu}, P_\mu)$)
\begin{equation}
\Delta _{0}(I_A )= I_A  \otimes 1 + 1 \otimes I_A  \, ,
\qquad
\Delta _{\theta}(I_A )= {\mathcal{F}}_\theta \circ \Delta _{0}(I_A )\circ
{\mathcal{F}}_\theta^{-1}
 \label{konlu14}
\end{equation}
and $(a\otimes b)\circ(c\otimes d) = ac\otimes bd$.

The dual canonical $\theta_{\mu\nu}$-deformed Poincar\'{e} group constructed
some years ago \cite{konlu10} and reconstructed recently
 \cite{konlu13,konlu14} looks as follows
\begin{eqnarray}
&&\left[ \hat{a}^{\mu },\hat{a}^{\nu }\right] =-\frac{i}{\kappa^2}\,\theta ^{\rho \sigma }_{(0)}( \hat{\Lambda} _{\
\rho }^{\mu }\,\hat{\Lambda} _{\ \sigma }^{\nu }-\delta _{\ \rho }^{\mu }\,\delta
_{\ \sigma }^{\nu }) \,,
\cr\cr
&&[ \hat{\Lambda} _{\ \tau }^{\mu },\hat{\Lambda} _{\ \rho }^{\nu }] =[
\hat{a}^{\mu },\hat{\Lambda} _{\ \rho }^{\nu }] =0\,,
 \label{konlu15}
\end{eqnarray}
with the coproducts remaining classical.

The twist \eqref{konlu12} provides an example of "soft" Abelian deformation,
with the classical $r$-matrix

\begin{equation}
r=\frac{1}{2\kappa ^{2}}\,\theta ^{\mu \nu }_{(0)}\,P_{\mu }\wedge P_{\nu }\,,
\label{konlu16}
\end{equation}
having the Abelian carrier algebra ($[P_\mu, P_\nu]=0$).

The twist \eqref{konlu16} determines the noncommutative structure
of $\theta_{\mu\nu}$-deformed Minkowski space ${\mathcal{M}}^{(\theta)}_4$.
In accordance with general framework for twisted quantum symmetries
\cite{konlu15,konlu16,konlu11,konlu12} the coproduct
formulae \eqref{konlu13} enter into the definition of noncommutative
associative *-product for the functions  $f,g \in {\mathcal{M}}^{(\theta)}_4$
\begin{equation}
f\hat{(x)}\star g\hat{(x)}:=\omega _{\theta}\left( f\hat{(x)}
\otimes g\hat{(x)}\right) =
\omega \left(
 \mathcal{F}^{-1}_{\theta}
  \circ f \hat{(x)}
  \otimes g \hat{(x)} \right) \,.
\label{konlu17}
\end{equation}
In such a description the quantum Minkowski space ${\mathcal{M}}^{(\theta)}_4$ is
a quantum representation (a Hopf module) of canonical $\theta_{\mu\nu}$-deformed
Poincar\'{e} algebra. The relation \eqref{konlu17} applied to
$f\hat{(x)}= \hat{x}_\mu$ and $g\hat{(x)}= \hat{x}_\nu$
provides the formula

\begin{equation}
\hat{x}_\mu \ast \hat{x}_\nu  = \frac{i}{2}
\theta^{(0)}_{\mu\nu} \,,
\label{konlu18}
\end{equation}
and leads to the relation \eqref{konlu8}. Interestingly enough,
the relation

\begin{equation}
[{x}_\mu, {x}_\nu ]_\ast   \equiv
{x}_\mu \ast {x}_\nu - {x}_\nu \ast {x}_\mu = i \theta_{\mu\nu}^{(0)} \,,
\label{konlu19}
\end{equation}
is covariant under the Hopf-algebraic action of twisted Poincar\'{e} algebra
generators. Using the general formula for the action of canonical quantum symmetry
generators
\begin{equation}
I_A \triangleright \omega_\theta (f(\hat{x})\otimes g(\hat{x}))=
\omega_\theta (\Delta_\theta (I_A) \circ
(f(\hat{x}) \otimes g(\hat{x})
)
\label{konlu20}
\end{equation}
and choosing $I_A \equiv M_{\mu\nu}$, one obtains
\begin{equation}
M_{\mu\nu}\triangleright ([x_\mu, x_\nu]_\ast) - i \theta_{\mu\nu})= 0 \, .
\label{konlu21}
\end{equation}
The algebraic multiplication formula \eqref{konlu17}
after the use of classical realizations
\begin{equation}
P_{\mu }\triangleright f(x)=i\partial _{\mu }f(x)\,,\qquad \qquad M_{\mu \nu } \triangleright
f(x)=i\left( x_{\mu }\partial _{\nu }-x_{\nu }\partial _{\mu }\right)
f(x)\,.  \label{konlu23}
\end{equation}
 can be represented on
commutative Minkowski space as the Moyal-Weyl star product
\begin{equation}
 f(x)\ast g(x) = \omega  (e^{\frac{i}{2\kappa^2} \theta^{(0)}_{\mu\nu}
  \partial^\mu \wedge \partial^\nu} \circ(  f(x) \otimes g(x)))
  \label{konlu22}
\end{equation}

The product  \eqref{konlu22} has been used recently quite often in order to
describe the effects of the noncommutativity of space-time coordinates
in classical and quantum field theory (see e.g. \cite{konlu17,konlu18}).

\section{Standard $\kappa$-Deformation: The Modification of High
 Energy Lorentz Boosts and Quantum Time}

The next class of quantum deformations is obtained by assuming that
the linear term  in \eqref{konlu11a} is nonvanishing. In order to
obtain the classical  nonrelativistic physics  we postulate
\begin{equation}
[\hat{x}_i , \hat{x}_j] = 0 \, .
  \label{konlu24}
\end{equation}
The only remaining $O(3)$-covariant deformed relation is the one
proposed firstly in the $\kappa$-deformed framework of Einstein's
symmetries \cite{konlu19,konlu20,konlu9}

\begin{equation}
[\hat{x}_0 , \hat{x}_i] = \frac{i}{\kappa} \hat{x}_i \, .
  \label{konlu25}
\end{equation}
We see from the relations \eqref{konlu25}--\eqref{konlu26}
that the space coordinates remain classical, but the time variable becomes quantum.
The  $\kappa$-deformed relativistic symmetries preserving the relations
\eqref{konlu24}--\eqref{konlu25} are generated by the following
 $\kappa$-deformed  Poincar\'{e}-Hopf algebra\footnote{We provide the formulae
 in so-called bicrossproduct basis, with classical Lorentz algebra}

a) algebraic sector

The only modified classical relation is between boosts (we put $c=1$ i.e.
$N_i=M_{i0}$)  and three momenta $P_j$:
\begin{equation}
[N_i, P_j] = i \delta_{ij} [\frac{\kappa}{2}(1-e^{- \frac{2P_0}{\kappa}})
+ \frac{1}{2\kappa} \, \overrightarrow{P}^2]
+ \frac{1}{\kappa} \, P_i P_j
\, ,
  \label{konlu26}
\end{equation}

b) coalgebra sector

We obtain the following set of coproducts for $N_i$ and $P_i$, satisfying the
relation \eqref{konlu26}

\begin{eqnarray}
\label{konlu27}
\Delta \, P_i & = & P_i \otimes 1 + e^{- \frac{P_0}{\kappa}} \otimes P_i \, ,
\cr\cr
\Delta \, N_i & = & N_i \otimes 1 + e^{- \frac{P_0}{\kappa}} \otimes N_i
+ \frac{1}{\kappa} \, \epsilon_{ijk} \, P_j \otimes M_k
\, .
\end{eqnarray}
Remaining coproducts for $P_0$ and $M_i$ are primitive, i.e.
\begin{eqnarray}
\label{konlu28}
\Delta \, P_0 & = & P_0 \otimes 1 + 1 \otimes P_0 \, ,
\cr\cr
\Delta \, M_i & = & M_i \otimes 1 + 1 \otimes M_i
\, ,
\end{eqnarray}

c) antipodes

The classical value $S(I_A) = -I_A$ is  modified for
the generators  $(P_i, N_i)$

\begin{eqnarray}
\label{konlu29}
S(P_i) &= & - e^{\frac{P_0}{\kappa}} \, P_i
\cr\cr
S(N_i) &= & - e^{\frac{P_0}{\kappa}} \, N_i
+ \frac{1}{\kappa}\, \epsilon_{ijk} \, e^{\frac{P_0}{\kappa}}\, P_j \, M_k
\, .
\end{eqnarray}

Using the formulae \eqref{konlu26}--\eqref{konlu29} one can calculate
the $\kappa$-deformation of finite Lorentz transformations of the
fourmomenta. The classical Lorentz boost transformations
\begin{equation}
\label{konlu30}
P_\rho (\alpha_i) = e^{\alpha_i N_i} \
P_\rho \,  e^{- \alpha_i N_i}
\, ,
\end{equation}
in deformed case are generalized as follows
\begin{equation}
\label{konlu31}
P_\rho (\alpha_i) = {\rm ad}_{e^{\alpha_i N_i}}\, P_\rho
= \sum\limits^{\infty}_{n=0} \frac{\alpha_{i1}\ldots \alpha_{in} }{n!}
\left(
{\rm ad}_{{N_{i1}}} \ldots
 ( {\rm ad}_{{N_{in}}} \, P_\rho )
\right)
\, ,
\end{equation}
where the quantum adjoint action is defined by the formula

\begin{equation}
\label{konlu32}
{\rm ad}_Y \, X = Y_{(1)} \, X \, S(Y_{(2)}) \, .
\end{equation} 
where  $\Delta(Y) = Y_{(1)}\otimes Y_{(2)}$.
 Choosing  in \eqref{konlu32} $Y=N_i$ and $X=F(P_i, P_0)$ one obtains after
using \eqref{konlu26} and  \eqref{konlu29} (see e.g. \cite{konlu21})

\begin{equation}
\label{konlu34}
{\rm ad}_{N_i} \, F(P_i, P_0) = [N_i, F(P_i , P_0)]\, .
\end{equation}
If we use  \eqref{konlu34} in  \eqref{konlu31} we see that the
formula  \eqref{konlu31} takes the standard form  \eqref{konlu30}.
Choosing $\overrightarrow{\alpha}=(\alpha, 0,0)$ one obtains the
differential equation for $\kappa$-deformed Lorentz transformations
of $P_\rho$ in ($p_1, p_0$) plane

\begin{equation}
\label{konlu35}
\frac{dP_\rho(\alpha)}{d\alpha} = [N_1, P_\rho] \, .
\end{equation}
Substituting in  \eqref{konlu35} the rhs of  \eqref{konlu26} one gets the
nonlinear equation for $P_\rho$, which has been solved in bicrossproduct
basis firstly in \cite{konlu22} (the solution for arbitrary vector
 $\overrightarrow{\alpha} = \alpha \overrightarrow{n}$ ($\overrightarrow{n}^2 =1$)
 has been given in \cite{konlu23}).

 The $\kappa$-deformed Minkowski space  \eqref{konlu24}-- \eqref{konlu25}
  as the Hopf-algebra module of the $\kappa$-deformed Poincar\'{e}
  algebra has been firstly considered in \cite{konlu20}. Using

  \begin{equation}
  \label{konlu36}
  I_A \triangleright \omega(f\otimes g) =
  \omega(I_{A_{(1)}} \, \triangleright \,
  \, f \otimes
  I_{A_{(2)}} \, \triangleright \,
\,   g )\, ,
  \end{equation}
and the classical form of the actions
\begin{eqnarray}
\label{konlu37}
M_i \triangleright \hat{x}_j = \epsilon_{ijk} \hat{x}_k \, ,
&\qquad& M_i \triangleright \hat{x}_0 = 0 \, ,
\cr\cr
N_i \triangleright \hat{x}_j = -\delta_{ijk} \hat{x}_0 \, ,
&\qquad& N_i \triangleright \hat{x}_0 = -x_i \, ,
\end{eqnarray}
one gets
\begin{eqnarray}
\label{konlu38}
N_i \triangleright \hat{x}_0 \hat{x}_j &=& \delta_{ij} \hat{x}_0^2
- \hat{x}_i \hat{x}_j + \frac{1}{\kappa} \delta_{ij} \hat{x}_0 \, ,
\cr\cr
N_i \triangleright \hat{x}_j \hat{x}_0 &= &-\delta_{ij} \hat{x}_0^2
- \hat{x}_i \hat{x}_j \, ,
\end{eqnarray}
and subsequently
\begin{equation}
\label{konlu39}
N_i \triangleright ([\hat{x}_0, \hat{x}_0]- \frac{1}{\kappa^2} x_i) = 0\, .
\end{equation}
Similarly one can show that $N_i \triangleright ([\hat{x}_i, \hat{x}_j]=0$.

The multiplication leading to the formulae \eqref{konlu24}--\eqref{konlu25}
can be represented by a particular CBH *-product (CBH$\equiv$Cambell-Baker-Haussdorf)
 on commutative Minkowski space $x_\mu$. From
 \eqref{konlu24}--\eqref{konlu25} follows that (see \cite{konlu24})
 \begin{equation}
 \label{konlu40}
 e^{ip^4 \hat{x}_\mu}\cdot e^{ik^\mu\hat{x}_\mu}=
 e^{i r^\mu(p,k)\hat{x} _\mu} \, ,
\end{equation}
where
 \begin{equation}
 \label{konlu41}
 r^0 = p^0 +k^0  \, , \qquad
 r^i = \frac{f_\kappa (p^0) e^{\frac{k^0}{\kappa}p^i + f_\kappa (k^0) k^i}}
 {f_\kappa(p^0 +k^0)} \, ,
 \end{equation}
and
 $f_\kappa(\alpha) = \frac{\kappa}{\alpha}(1-e^{-\frac{\alpha}{\kappa}})
 = 1 - \frac{1}{2} \frac{\alpha}{\kappa} + \mathcal{O}(\frac{1}{\kappa^2})$.

The CBS star product representing the multiplication rule \eqref{konlu40}
can be described by the formula
\begin{equation}
\label{konlu42}
f(x) \ast g(x) = f(\frac{1}{i} \frac{\partial}{\partial p^\kappa})
g(\frac{1}{i} \frac{\partial}{\partial k^\kappa})
e^{i r^4(p,k) x_\mu}\, .
\end{equation}
In particular we obtain from \eqref{konlu42}
\begin{eqnarray}
\label{konlu43}
x_0 \ast x_i & = & x_0 x_i + \frac{i}{2\kappa} x_i \, ,
\cr\cr
x_i \ast x_0 & = & x_0 x_i -  \frac{i}{2\kappa} x_i \, ,
\end{eqnarray}
i.e. we reproduce the relation \eqref{konlu25}.

After the discovery of $\kappa$-deformation
of the Poincar\'{e} symmetries in 1991 there was found in 1995
the light-cone deformation of Poincar\'{e} algebra \cite{konlu25}, with one
of the light-cone directions quantized, and the relations
\eqref{konlu24}--\eqref{konlu25} replaced by the following ones
($r,s=1,2$)
\begin{eqnarray}
\label{konlu44}
&[\hat{x}_-, \hat{x}_r ]  =  \frac{1}{\kappa}\,  x_r \, ,
\qquad
[\hat{x}_-, \hat{x}_+ ]  =  \frac{1}{\kappa} \hat{x}_-\, ,
\cr\cr
& [\hat{x}_r, \hat{x}_s ]  =
[\hat{x}_+, \hat{x}_r ] = 0 \, ,
\end{eqnarray}
where $x_{\pm} = x_0 \pm x_3$. In 1996 both relations
\eqref{konlu24}--\eqref{konlu25} and \eqref{konlu44} were obtained as special
cases of generalized $\kappa$-deformation of Poincar\'{e}
symmetries \cite{konlu26,konlu17} with the quantized noncommutative direction
$a_\mu \hat{x}^\mu$ in deformed Minkowski space. We get the following
$a_\mu$-dependent noncommutativity structure of quantum space-time
\begin{equation}
\label{konlu45}
[\hat{x}_\mu, \hat{x}_\nu] = \frac{i}{\kappa} (a_\mu \, \hat{x}_\nu -
a_\nu \, \hat{x}_\mu)\, .
\end{equation}
It follows from \cite{konlu26}--\cite{konlu28} that only if
$a^2_\mu=0$, i.e. when the light-cone direction in Minkowski space is quantized,
 the corresponding deformation of Poincar\'{e} symmetries can be
 represented by the twist factor,
 which modifies by the similarity transformation the
 classical coproduct (see \eqref{konlu14}).
 In such a case only the coalgebraic sector is modified and
   we deal with twisted quantum relativistic
 symmetries.

 It should be added that $\kappa$-deformations were further employed from 2000
 as a mathematical structure describing so-called doubly special relativistic
  (DSR) theories  (see e.g. \cite{konlu29}--\cite{konlu31}).
  Basic aspects of the relation between the DSR framework and the
  $\kappa$-deformation of Poincar\'{e} symmetries were discussed by the
  present author in \cite{konlu23,konlu32}.

  \section{Outlook}
  The idea of noncommutative space-time can be traced back to Heisenberg
  who first indicated that a space-time uncertainty relation might be
  useful for the removal of infinities in renormalization procedure.
  Unfortunately till present time the idea of compensating the divergent
  terms (see e.g. \cite{konlu33}) in noncommutative field theory has not
  been successful. Also other ways of justifying the modifications of classical
  relativistic symmetries, based on the observation of modified kinematics
  in cosmic ray physics (see e.g. \cite{konlu34}) after a period of hopeful
  uncertainties in the interpretation of experimental data
    at present are rather not confirmed.

  Important theoretical arguments for noncommutativity of space-time are
  based on quantum gravity and quantized string theory. They indicate that
  the implementation of quantum symmetries and quantum modification of
  relativistic symmetries should be a necessary step for the
  description of the distances comparable with the Planck length $l_P$ ($l_P\simeq
   10^{-33}$cm).
  Also the role of noncommutative geometry in the ultimate unification
  of all interactions - the $M$-theory - appears to be important and
   should be further explored.

  Now however the main challenge is to find among experimental data in high energy
  physics and astrophysics the ones which indicate the violation of
  the postulates of Einstein special relativity theory (e.g. constant velocity
   of light, classical conservation law of fourmomenta, classical energy
   momentum dispersion relation, isotropy of "empty" space-time etc.).
 One can conjecture that these possible classical symmetry breaking effects
    can be recast into a new quantum relativistic symmetry invariance.

\begin{theacknowledgments}
The author would like to thank the organizers of the Albert Einstein Century
International Conference for warm hospitality and acknowledge the financial
support by KBN grant 1P03B01828.
\end{theacknowledgments}

\end{document}